\documentclass[sn-mathphys,Numbered]{sn-jnl}

\usepackage{graphicx}%
\usepackage{multirow}%
\usepackage{amsmath,amssymb,amsfonts}%
\usepackage{amsthm}%
\usepackage{mathrsfs}%
\usepackage[title]{appendix}%
\usepackage{xcolor}%
\usepackage{textcomp}%
\usepackage{manyfoot}%
\usepackage{booktabs}%
\usepackage{algorithm}%
\usepackage{algorithmicx}%
\usepackage{algpseudocode}%
\usepackage{listings}%

\theoremstyle{thmstyleone}%
\theoremstyle{thmstyletwo}%

\theoremstyle{thmstylethree}%

\begin{document}

\title[Tunable CD]{A Programmable Wafer-scale Chiroptical Heterostructure of Twisted Aligned Carbon Nanotubes and Phase Change Materials}


\author[1]{\fnm{Jichao} \sur{Fan}}

\author[1]{\fnm{Ruiyang} \sur{Chen}}

\author[1]{\fnm{Minhan} \sur{Lou}}

\author[1]{\fnm{Haoyu} \sur{Xie}}

\author[2]{\fnm{Nina} \sur{Hong}}

\author[1]{\fnm{Yingheng} \sur{Tang}}

\author*[1]{\fnm{Weilu} \sur{Gao}}\email{weilu.gao@utah.edu}

\affil[1]{Department of Electrical and Computer Engineering, The University of Utah, Salt Lake City, UT, USA}

\affil[2]{J.A. Woollam Co., Inc., Lincoln, NE, USA}


\abstract{The ability to design and dynamically control chiroptical responses in solid-state matter at wafer scale enables new opportunities in various areas. Here we present a full stack of computer-aided designs and experimental implementations of a dynamically programmable, unified, scalable chiroptical heterostructure containing twisted aligned one-dimensional (1D) carbon nanotubes (CNTs) and non-volatile phase change materials (PCMs). We develop a software infrastructure based on high-performance machine learning frameworks, including differentiable programming and derivative-free optimization, to efficiently optimize the tunability of both excitonic reciprocal and linear-anisotropy-induced nonreciprocal circular dichroism (CD) responses. We experimentally implement designed heterostructures with wafer-scale self-assembled aligned CNTs and deposited PCMs. We dynamically program reciprocal and nonreciprocal CD responses by inducing phase transitions of PCMs, and nonreciprocal responses display polarity reversal of CD upon sample flipping in broadband spectral ranges. All experimental results agree with simulations. Further, we demonstrate that the vertical dimension of heterostructure is scalable with the number of stacking layers and aligned CNTs play dual roles – the layer to produce CD responses and the Joule heating electrode to electrically program PCMs. This heterostructure platform is versatile and expandable to a library of 1D nanomaterials and electro-optic materials for exploring novel chiral phenomena and photonic and optoelectronic devices.}

\maketitle

\section*{Introduction}

Manipulating chiroptical responses in solid-state materials and their photonic and optoelectronic devices has enabled various applications, including sensing~\cite{LiuEtAl2023NRC,YooEtAl2019N,WarningEtAl2021N}, imaging~\cite{ZhanEtAl2021AM,KhaliqEtAl2022AOM}, neuromorphic computing~\cite{DanEtAl2024N}, and light-driven synthesis~\cite{KimEtAl2019JACS,GenetEtAl2022P}. In particular, the interaction of circularly polarized light with quantum materials has profound implications on quantum photonic applications~\cite{LodahlEtAl2017N,HubenerEtAl2021NM,ChenEtAl2021N,AielloEtAl2022N,LiEtAl2023NM}. Chiroptical responses, such as circular dichroism (CD) that is the differential attenuation between the left- and right-handed circularly polarized (LCP and RCP) light, generally have two prominent contributions with distinct features~\cite{AlbanoEtAl2020CR,AlbanoEtAl2022C}. The first one is an intrinsic isotropic reciprocal response due to molecular or structure-induced chirality, which is invariant upon sample orientation. The second one is a linear anisotropy-induced nonreciprocal response, leading to opposite handednesses from opposite sides of the same sample. Leveraging both contributions of chiroptical responses can enrich functionalities, improve efficiencies, and reduce costs of chiral photonic and optoelectronic devices~\cite{FurlanEtAl2024NP}.

Recently, twisted stacks of self-assembled aligned 1D nanomaterials, such as carbon nanotubes (CNTs), have emerged as a high-performance chiral photonic material platform~\cite{DoumaniEtAl2023NC}. For example, a 3-layer twisted stack of aligned CNTs has demonstrated the highest CD response compared to all other material and metamaterial platforms in the deep ultraviolet spectral range, because of the strong room-temperature quantum-confinement effects in CNTs~\cite{DresselhausEtAl2001,AvourisEtAl2008NP} compared to natural molecules and self-assembled conventional metallic or dielectric materials without quantum-relevant properties~\cite{LvEtAl2022NRC}. Further, the optical properties of 1D nanomaterials can be structurally and dynamically tuned in a broadband spectral range from ultraviolet to infrared frequencies. In addition, in contrast to engineered metamaterials, whose top-down manufacturing requires sophisticated nanofabrication facilities and processes to create artificial symmetry-breaking structures and is challenging to scale up~\cite{KaschkeEtAl2016N,WangEtAl2016N,MunEtAl2020LSA}, bottom-up self-assembly of 1D nanomaterials is low-cost and at wafer scale~\cite{ZhaoEtAl2024S}. Although hybridizing metamaterials with materials or structures whose optical properties or geometries can be electrically, thermally, or mechanically controlled enables dynamic tunability~\cite{YinEtAl2015NL,KimEtAl2017SA,KindnessEtAl2020AOM,KwonEtAl2021P,RodriguesEtAl2022JN}, this hybridization further increases manufacturing complexity and the design of dynamically tunable metametarials using full-wave simulations is time-consuming and computation-intensive~\cite{JiangEtAl2020NRM}. Moreover, nonreciprocal chiroptical responses are occasionally reported in a few solid-state films of organic materials without systematic designs~\cite{AlbanoEtAl2022C,FurlanEtAl2024NP}, and are largely ignored or missing in metamaterial reports. Hence, a fully programmable, unified, wafer-scale chiroptical platform, whose reciprocal and nonreciprocal chiroptical responses can both be efficiently designed through programming and can be electrically tuned and dynamically programmed, is lacking. 


Here, we present such a platform that is a heterostructure of twisted aligned 1D nanomaterials and non-volatile chalcogenide phase change materials (PCMs). The demonstrated 1D nanomaterial and PCM are CNT and germanium-antimony-tellurium (Ge$_2$Sb$_2$Te$_5$ or GST), respectively. The material phases of PCMs can be fast electrically programmed to be crystalline or amorphous using a short electrical pulse with a substantial dielectric function modulation in broadband spectral ranges~\cite{WuttigEtAl2017NP,ZhangEtAl2019NC}. Further, the phases can be preserved after programming without external stimulus for $>10$ years with zero static energy consumption. We implemented simulation software based on machine learning frameworks, including graphics processing unit (GPU)-powered differentiable programming enabled by gradient backpropagation algorithm and derivative-free Bayesian optimization, to efficiently design and optimize the dynamic tunable ranges of reciprocal and nonreciprocal CD responses in the heterostructure with experimentally obtained optical constants of materials. Further, we experimentally implemented designed heterostructures through wafer-scale low-cost self-assembly of aligned CNTs and wafer-scale deposition of GST and dielectric films, and experimentally observed strong dynamic tunability. In particular, we demonstrated tunable nonreciprocal CD responses, which displayed opposite signs or polarity reversal of measured CD signals in a broadband visible range when opposite sides of the same heterostructure were probed. All measured spectra agreed with the simulations. In addition, we demonstrated that not only the lateral dimension of the heterostructure was at wafer scale, but also the vertical dimension was scalable with a large number of stacking layers to enhance CD responses. Moreover, we showed that aligned CNTs in the heterostructure played dual roles -- the active material to produce CD responses and the Joule heating electrode material to electrically program GST phases. The demonstrated heterostructure architecture is versatile and can be extended to incorporate a wide range of 1D nanomaterials, such as transition metal dichalcogenides nanotubes~\cite{YomogidaEtAl2018O}, boron nitride nanotubes~\cite{XuEtAl2022CEJ}, nanotube heterostructures~\cite{XiangEtAl2020S}, and encapsulated CNTs with 1D atomic chains~\cite{BalandinEtAl2022MT}, and different PCMs and electro-optic materials~\cite{Benea-ChelmusEtAl2021NC,PrabhathanEtAl2023I,BoesEtAl2023S}, such as antimony sulfide (Sb$_2$S$_3$), antimony selenide (Sb$_2$Se$_3$), lithium niobate (LiNO$_3$) and electro-optic organic materials. This versatile chiroptical platform will provide opportunities for novel chiral photonic and optoelectronic devices, such as chiral quantum light emitters~\cite{LiEtAl2023NM}, and provide a new platform for exploring optical and non-optical chiral phenomena, such as chirality-induced spin selectivity effects~\cite{BloomEtAl2024CR}.



\section*{Results}

\subsection*{Programmable heterostructure architecture}
\begin{figure}[hbt]
    \centering
    \includegraphics[width=1.0\textwidth]{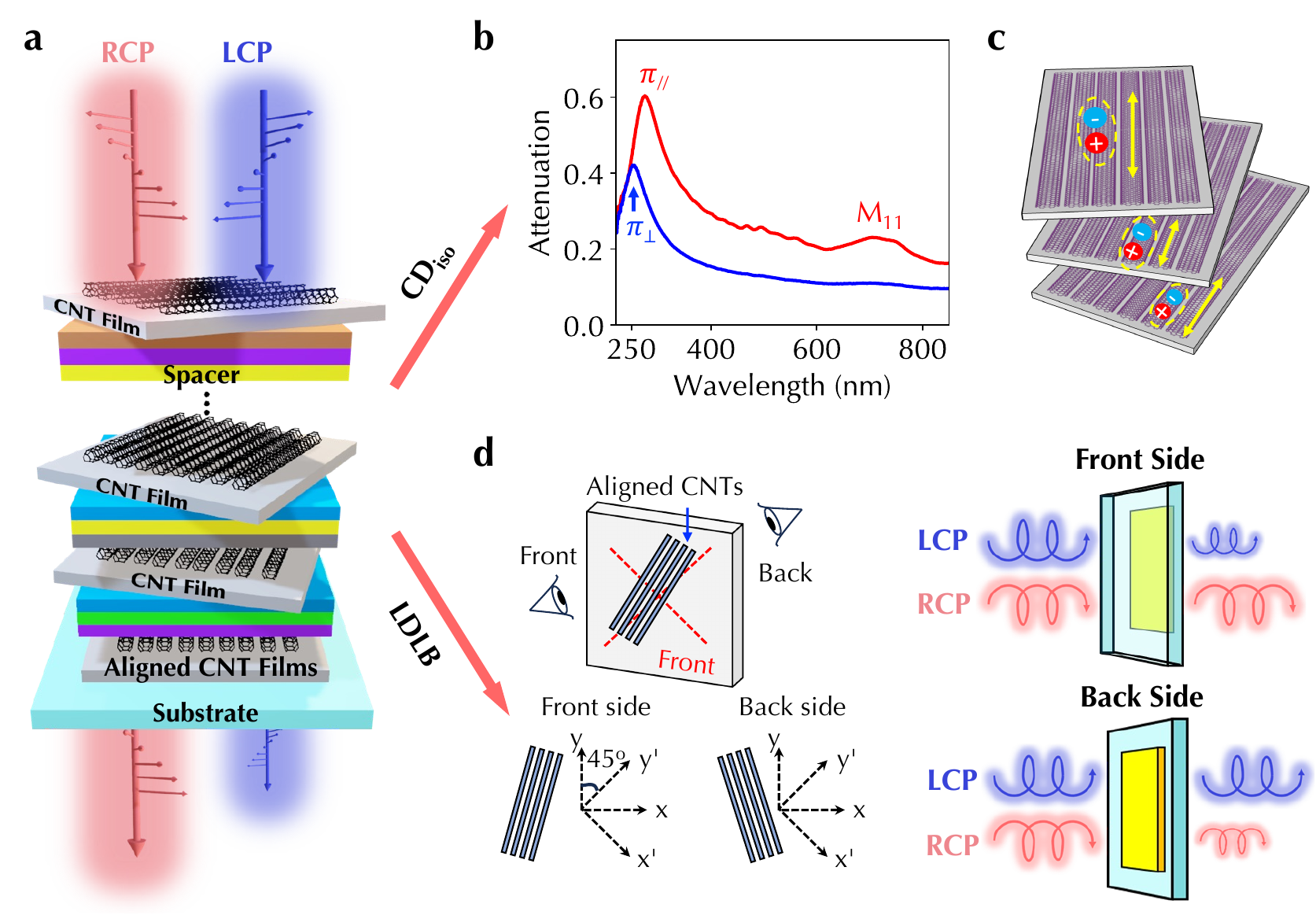}
    \caption{\textbf{Programmable heterostructure architecture and chiroptical responses}. (a)~Illustration of programmable chiroptical heterostructure containing anisotropic aligned CNTs and isotropic PCMs and dielectrics. (b)~Linear-polarization-dependent absorption spectra of an aligned CNT film prepared using the SAVF process. The red (blue) line for parallel (perpendicular) polarization. (c)~Isotorpic reciprocal CD$_\mathrm{iso}$ response in the heterostructure originating from the helical coupling of anisotropic 1D excitons in aligned CNTs. (d)~Nonreciprocal LDLB response in the heterostructure originating from the interference of linear anisotropy, which displays opposite handednesses from the front and back sides of the same sample.}
    \label{fig:fig1}
\end{figure}

Figure~\ref{fig:fig1}a schematically illustrates the programmable heterostructure architecture containing multiple layers of anisotropic aligned CNTs and other isotropic materials, including PCMs and dielectrics. The quantum confinement along CNT circumferences induces the formation of subbands and excitonic interband transitions across subbands. These transitions span broadband spectral ranges and are dependent on the atomic structures of CNTs~\cite{DresselhausEtAl2001,AvourisEtAl2008NP}. Further, the 1D geometry of CNTs leads to anisotropic excitonic electric dipoles and optical absorption under the excitation of linearly polarized light. We used a shaking-assisted vacuum filtration (SAVF) process to prepare aligned CNT films and measured their linear-polarization-dependent absorption spectra. The linear shaking occurring during the vacuum filtration dictated the alignment direction of obtained aligned CNT films; see Methods and Supplementary Fig.~1 for more details. The deterministic control of CNT alignment direction in the SAVF process is advantageous over conventional vacuum filtration without direction control and facilitates the scalable stacking of multiple layers in the heterostructure. As shown in Fig.~\ref{fig:fig1}b, when the polarization of incident light is parallel to the CNT axis, strong absorption occurs through M$_{11}$ transition in CNTs at 730\,nm and the $M$ point transition (labeled `$\pi_{\parallel}$') at 282\,nm~\cite{MakEtAl2011PRL}. The absorption resonance wavelength is dependent on the diameters and atomic structures of CNTs~\cite{LiuEtAl2011NC}. In contrast, different $M$ point transition transition (labeled `$\pi_{\perp}$') occurs at 255\,nm due to optical selection rules~\cite{TakagiEtAl2009PR} and the M$_{11}$ transition is suppressed because of the depolarization effect~\cite{YanagiEtAl2018NC}. The helical twist of anisotropic excitonic dipoles and electromagnetic coupling between them, which essentially correspond to the physical picture of helically coupled electrical dipoles in microscopic chiral molecules~\cite{NordenEtAl2010}, produce reciprocal CD response (CD$_\mathrm{iso}$); see Fig.~\ref{fig:fig1}c. The CD$_\mathrm{iso}$ response is of intrinsic excitonic nature, isotropic, and independent of sample orientation. The incorporation of PCMs and other isotropic materials in the heterostructure can control the electromagnetic coupling between dipoles to dynamically tune and optimize CD$_\mathrm{iso}$ response.

In addition, the other contributor to observed CD response originates from the interference of linear dichroism (LD) and linear birefringence (LB) of aligned CNTs. The LDLB-induced CD response is expressed as 0.5(LD$^{'}$LB $-$ LDLB$^{'}$), where LD and LB are measured in arbitrarily defined $x$-$y$ axes and LD$^{'}$ and LB$^{'}$ are measured along the bisectors of $x$-$y$ axes (i.e., with a $+45^{\circ}$ rotation). Note that the LDLB response is not from excitons, independent of instrumental faults, and not an artifact~\cite{AlbanoEtAl2020CR,AlbanoEtAl2022C}. Figure~\ref{fig:fig1}d illustrates an aligned CNT film on a plane, where $x$-$y$ axes define LD and LB and $x^{'}$-$y^{'}$ axes define LD$^{'}$ and LB$^{'}$. When looking from the front or back side of the plane or flipping the sample, aligned CNTs become mirrored with LD and LB unaltered and LD$^{'}$ and LB$^{'}$ signs flipped. Hence, the sign of the LDLB response is also flipped and called nonreciprocal CD~\cite{AlbanoEtAl2020CR,AlbanoEtAl2022C}. If the LDLB response is much larger than isotropic CD$_\mathrm{iso}$, the measured CD spectra can display opposite signs when measuring from opposite sample sides; see Fig.~\ref{fig:fig1}d. The LDLB response is non-zero only when the main axes of LD and LB are not aligned with each other otherwise two terms in the LDLB expression are always canceled. However, the optical axes of maximum attenuation for LD and phase delay for LB of CNTs are both along the CNT axis and minimum ones are both perpendicular to the CNT axis, meaning LD and LB quantities are aligned. This suggests the LDLB response be zero in CNT-only architectures. The incorporation of PCMs and other isotropic materials in the heterostructure can induce and control multiple reflections between layers. Because LD (LB) is related to the imaginary (real) part of the CNT refractive index, the multiple reflections inside the same heterostructure can have different influences on LD and LB and their main axes are decoupled and different from the CNT axis. Hence, the heterostructure not only can generate non-zero LDLB responses but can dynamically tune and optimize LDLB responses.

\subsection*{Simulation framework}

\begin{figure}[hbt]
    \centering
    \includegraphics[width=1.0\textwidth]{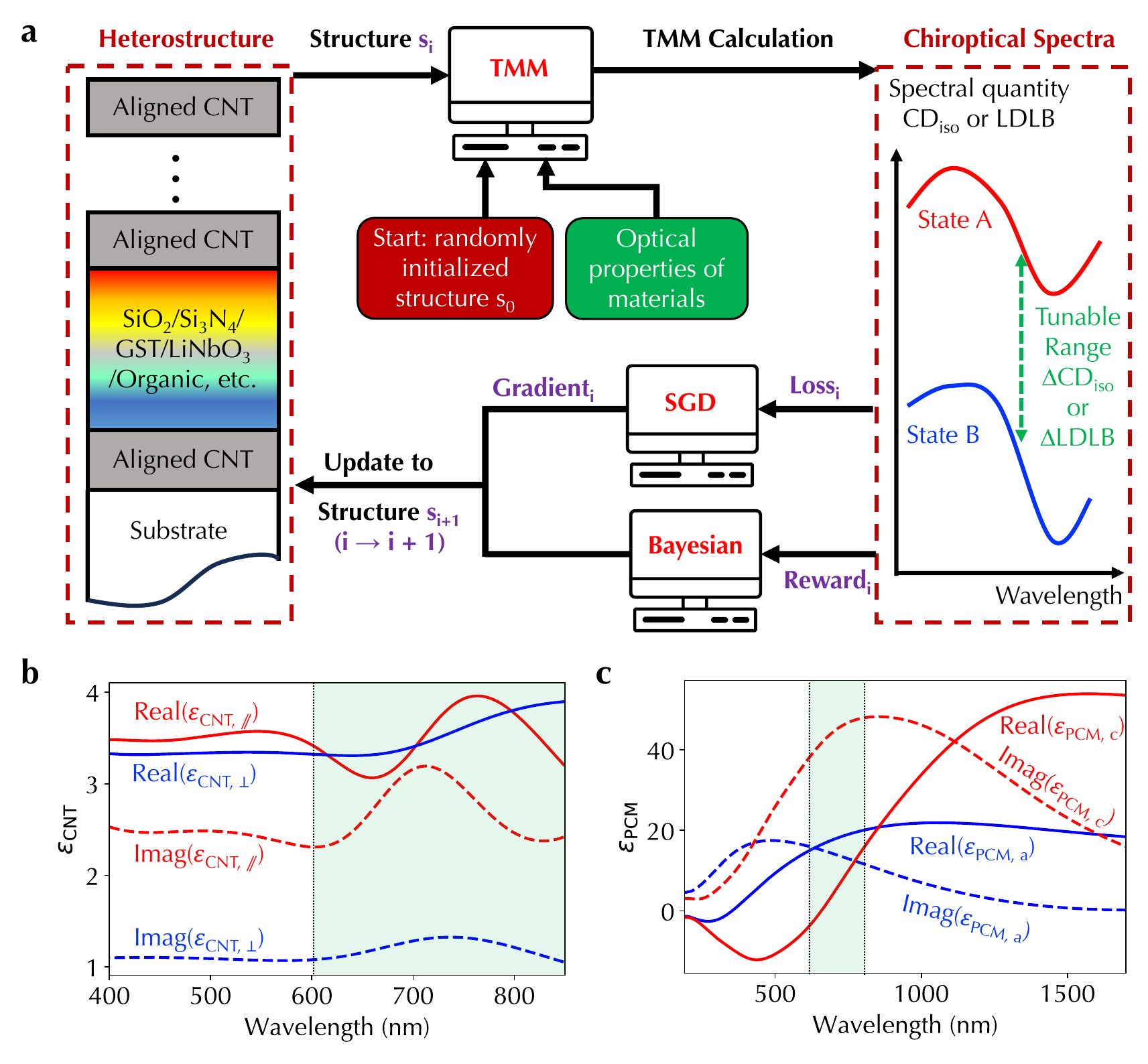}
    \caption{\textbf{Simulation framework based on machine learning frameworks}. (a)~Flowchart of designing the chiroptical heterostructure to achieve prominent dynamic tunability ranges of CD$_\textrm{iso}$ or LDLB ($\Delta$CD$_\textrm{iso}$ or $\Delta$LDLB) through differentiable programming enabled by the stochastic gradient descent (SGD) algorithm through the backpropagation algorithm in \texttt{PyTorch} and derivative-free Bayesian optimization algorithm. The framework takes input from experimentally determined complex-valued dielectric functions of (b)~an anisotropic aligned CNT film and (c)~a sputtered isotropic GST film. In (b), the red (blue) solid line is the real part of the dielectric function of the aligned CNT film along (perpendicular to) the CNT axis. In (c), the red (blue) solid line is the real part of the dielectric function of the GST film under the crystalline (amorphous) phase. The dashed lines are imaginary parts in both (b) and (c).}
    \label{fig:fig2}
\end{figure}

Figure~\ref{fig:fig2}a illustrates the simulation framework we developed to design the heterostructure to achieve prominent dynamic tunability of both CD$_\mathrm{iso}$ and LDLB responses. We developed a general $4\times4$ transfer matrix method (TMM) to obtain all transmission and reflection coefficients under different linear and circular polarizations; see Methods and Supplementary Fig.~2 for more details. We calculated these coefficients under four configurations of sample rotation and flipping to obtain CD$_\mathrm{iso}$ and LDLB responses; see Methods and Supplementary Fig.~3 for more details. Since all fundamental operations in TMM are matrix multiplications, we implemented the TMM framework using the \texttt{PyTorch} framework so that all calculations were GPU-accelerated in parallel. The TMM input included the dielectric functions of anisotropic and isotropic materials and a set of heterostructure parameters $s$, such as the choice of materials, layer thicknesses, and rotation angles between aligned CNTs. Figure~\ref{fig:fig2}b and \ref{fig:fig2}c display our experimentally determined complex-valued dielectric functions of aligned CNT films through multi-curve fittings and sputtered GST films under amorphous and crystalline phases through spectroscopic ellipsometry; see Methods and Supplementary Fig.~4 for more details. We focused on the spectra range covering the $M_{11}$ transition of CNTs because of the relatively low optical loss of GST at least in one phase in this range and the limitation of the spectrometer measurement range (shaded green areas in Figure~\ref{fig:fig2}b and \ref{fig:fig2}c). Note that although we focused on the non-volatile GST material for programming the heterostructure, the candidate materials can be broadly expanded to other electro-optic materials, such as electro-optic LiNbO$_3$ and organic materials. 

To design the heterostructure to achieve optimal tunable ranges of CD$_\mathrm{iso}$ and LDLB responses, a randomly initiated structure $s_0$ was input into the TMM to calculate CD$_\mathrm{iso}$ and LDLB responses when the dielectric functions of amorphous and crystalline GST were used. The difference of CD$_\mathrm{iso}$ (LDLB) spectra in the wavelength range of interest under two states of GST, denoted as $\Delta$CD$_\textrm{iso}$ ($\Delta$LDLB), was used to define optimization target functions. In principle, any other physical quantities calculated from transmission and reflection coefficients can become target functions, such as the dissymmetry $g$ factor defined as the ratio of CD$_\mathrm{iso}$ over the attenuation under unpolarized light. We developed two different categories of optimization algorithms. The first one is based on the stochastic gradient descent algorithm implemented using \texttt{PyTorch}-supported backpropagation algorithm. Since the backpropagation algorithm aims to minimize a target loss function, the loss function was defined as the most negative absolute value of $\Delta$CD$_\textrm{iso}$ or $\Delta$LDLB spectra. Based on the gradient calculated from the backpropagation algorithm, the heterostructure parameters were updated. Hence, in $i$-th iteration, the set of heterostructure parameters $s_i$ was updated to $s_{i+1}$ to decrease the loss function, which was equivalent to increase $\Delta$CD$_\textrm{iso}$ or $\Delta$LDLB. The optimal heterostructure was obtained after multiple iterations.

In addition to the gradient-based algorithm, we also demonstrated the feasibility of derivative-free algorithms to design the heterostructure. Note that although our developed TMM solver using \texttt{PyTorch} is differentiable, the derivative-free algorithm can be broadly applicable even when gradients are not available in TMM implementations. Specifically, we utilized the Bayesian optimization algorithm. In contrast to the gradient-based algorithm, the target of the Bayesian optimization algorithm is to maximize the reward function. Hence, we defined the largest absolute value of $\Delta$CD$_\textrm{iso}$ or $\Delta$LDLB spectra as the reward function. In each iteration, based on prior sets of structural parameters and reward functions, the algorithm determined the Bayesian posterior probability and a new set of structural parameters through sampling. After multiple iterations, the heterostructures that display $\Delta$CD$_\textrm{iso}$ or $\Delta$LDLB were highly probably sampled.

\subsection*{Experimental demonstration}

\begin{figure}[hbt]
    \centering
    \includegraphics[width=1.0\textwidth]{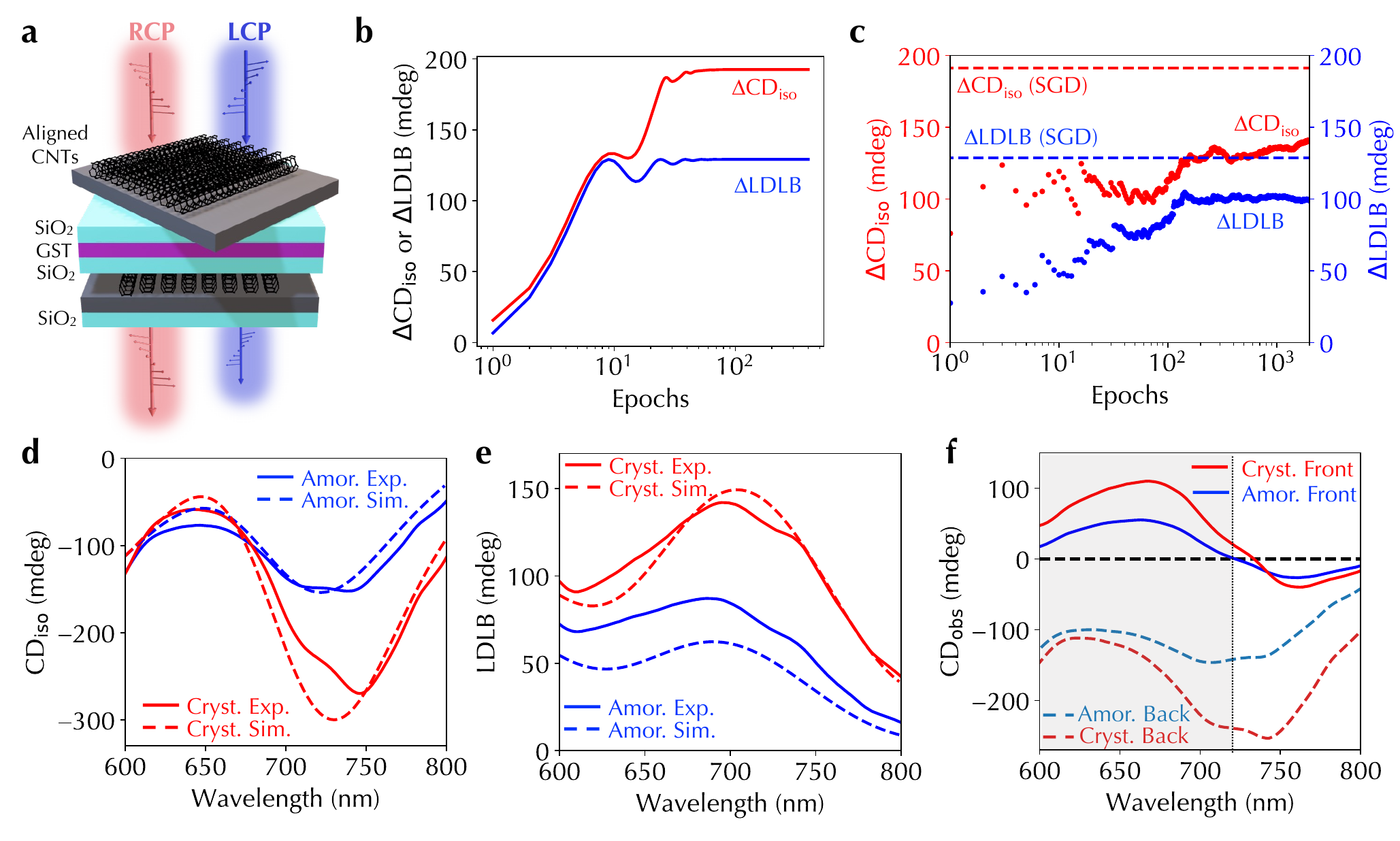}
    \caption{\textbf{Simulation and experimental implementation of a programmable heterostructure with two layers of aligned CNT films}. (a)~Illustration of the specific heterostructure containing two layers of twisted aligned CNT films. Training curves of $\Delta$CD$_\textrm{iso}$ (red lines in (b) or dots in (c)) and $\Delta$LDLB (blue lines in (b) or dots in (c)) using (b)~differentiable programming and (c)~derivative-free Bayesian optimization, respectively. Red (blue) dashed lines in (c) indicate optimized values for $\Delta$CD$_\textrm{iso}$ and $\Delta$LDLB from (b). Measured (solid lines) and simulation (dashed lines) spectra of (d)~CD$_\textrm{iso}$ and (e)~LDLB responses under GST crystalline (red lines) and amorphous (blue lines) phases, respectively. (f)~Observed CD spectra measured from the front (solid lines) and back (dashed lines) sides of the same heterostructure under GST crystalline (red lines) and amorphous (blue lines) phases. The gray shaded area indicates the polarity reversal range. }
    \label{fig:fig3}
\end{figure}

Figure~\ref{fig:fig3}a illustrates a specific heterostructure consisting of two layers of twisted aligned CNTs, two silicon oxide (SiO$_{2}$) layers, and one GST layer between SiO$_{2}$ layers to demonstrate the optimization using the simulation framework and experimental validation. Figure~\ref{fig:fig3}b and \ref{fig:fig3}c display training curves using the differentiable backpropagation algorithm and derivative-free Bayesian optimization algorithm, respectively, for optimizing $\Delta$CD$_\textrm{iso}$ and the $\Delta$LDLB; see Methods for more details. The twisted angle between aligned CNTs and thicknesses of SiO$_{2}$ and GST layers were structural parameters to be optimized. The aligned CNT thickness was assumed to be a constant and benchmarked by comparing the ultraviolet attenuation with a standard sample; see Methods for more details. For Bayesian optimization, Figure~\ref{fig:fig3}c shows the accumulated average of $\Delta$CD$_\textrm{iso}$ and $\Delta$LDLB. In comparison, the average values from the Bayesian optimization are lower than those from the backpropagation algorithm because of relatively wide parameter ranges. If parameter ranges were set narrower, the Bayesian optimization produced similar optimization values to the backpropagation algorithm; see Supplementary Fig.~5a. The obtained twisted angle between two aligned CNT layers was 45$^{\circ}$ in both algorithms (Supplementary Fig.~5b and Fig.~5c), which agrees with the optimized angle in CNT-only twisted stack~\cite{DoumaniEtAl2023NC}. 

We fabricated optimized heterostructures by transferring SAVF-prepared aligned CNT films and depositing GST and SiO$_{2}$ films using sputtering on fused silica substrates, which are transparent in the measurement spectra range (200\,nm -- 900\,nm); see Methods for more details. Similar to the simulation framework, we performed four CD measurements using a standard CD spectrometer with a compatible 3D-printed sample holder under four configurations of sample rotation and flipping to obtain CD$_\mathrm{iso}$ and LDLB responses; see Methods for more details. The as-deposited GST film was in the amorphous phase and was converted into the crystalline phase by heating the sample; see Methods, Supplementary Fig.~6, and Supplementary Video 1 for more details. Figure~\ref{fig:fig3}d and \ref{fig:fig3}e show measured CD$_\mathrm{iso}$ and LDLB spectra (solid lines) when the GST film was under amorphous and crystalline phases, which agreed with the simulation spectra (dashed lines) using the developed simulation framework. The maximum obtained $\Delta$CD$_\textrm{iso}$ was 123\,mdeg at 750\,nm wavelength and the maximum obtained $\Delta$LDLB was 60\,mdeg at 737\,nm wavelength. Broadband CD$_\mathrm{iso}$ and LDLB spectra under two GST phases are shown in Supplementary Fig.~7a and 7b. As shown in Supplementary Fig.~7c, the measured average absorption spectra of LCP and RCP light of the heterostructure under two GST phases (solid lines) also showed agreement with simulations (dashed lines). Further, we designed and optimized the heterostructure for the tunable range of $g$ factor using the backpropagation algorithm and experimentally measured spectra, which also agreed with simulation spectra (Supplementary Fig.~8). This demonstrates the broad applicability of the simulation framework and experimental implementation of the heterostructure. 

In particular, prominent LDLB responses were observed in Fig.~\ref{fig:fig3}e. For the same heterostructure, Fig.~\ref{fig:fig3}f and Supplementary Fig.~9 display measured CD spectra (CD$_\mathrm{obs}$) when the front and back sides of the heterostructure faced the direction of incident light; see Methods for more details. We observed a clear sign or polarity reversal of CD$_\mathrm{obs}$ in a broadband wavelength range of $424 - 721$\,nm for the GST amorphous phase and a wavelength range of $421 - 735$\,nm for the crystalline phase. As mentioned before, the LDLB response originates from the difference between intensity attenuation and phase delay extrema due to the interference of multiple reflections in the heterostructure. To confirm this, we further employed the developed simulation framework to calculate linear-polarization-dependent transmitted intensity and the phase delay between transmitted light and input light at a specific wavelength, as illustrated in Supplementary Fig.~10a. The zero degree was defined when the orientation of the first aligned CNT layer and the polarization direction were the same. Supplementary Fig.~10b and 10c display the calculated normalized attenuation and phase delay as a function of polarization angle at 630\,nm for two GST phases, showing clear shifts of extrema positions. While maximum phase delay positions occurred at $0^{\circ}$ under both GST phases, maximum attenuation positions occurred at $16.7^{\circ}$ and $13.3^{\circ}$, respectively. Note these values are also different from the orientation of second-layer aligned CNTs, which is $45^{\circ}$ under the coordination system defined in Supplementary Fig.~10a. We also experimentally measured attenuation at 630\,nm of the heterostructure as a function of polarization angle, confirming the shift, as shown in Supplementary Fig.~10d. 

\begin{figure}[hbt]
    \centering
    \includegraphics[width=0.9\textwidth]{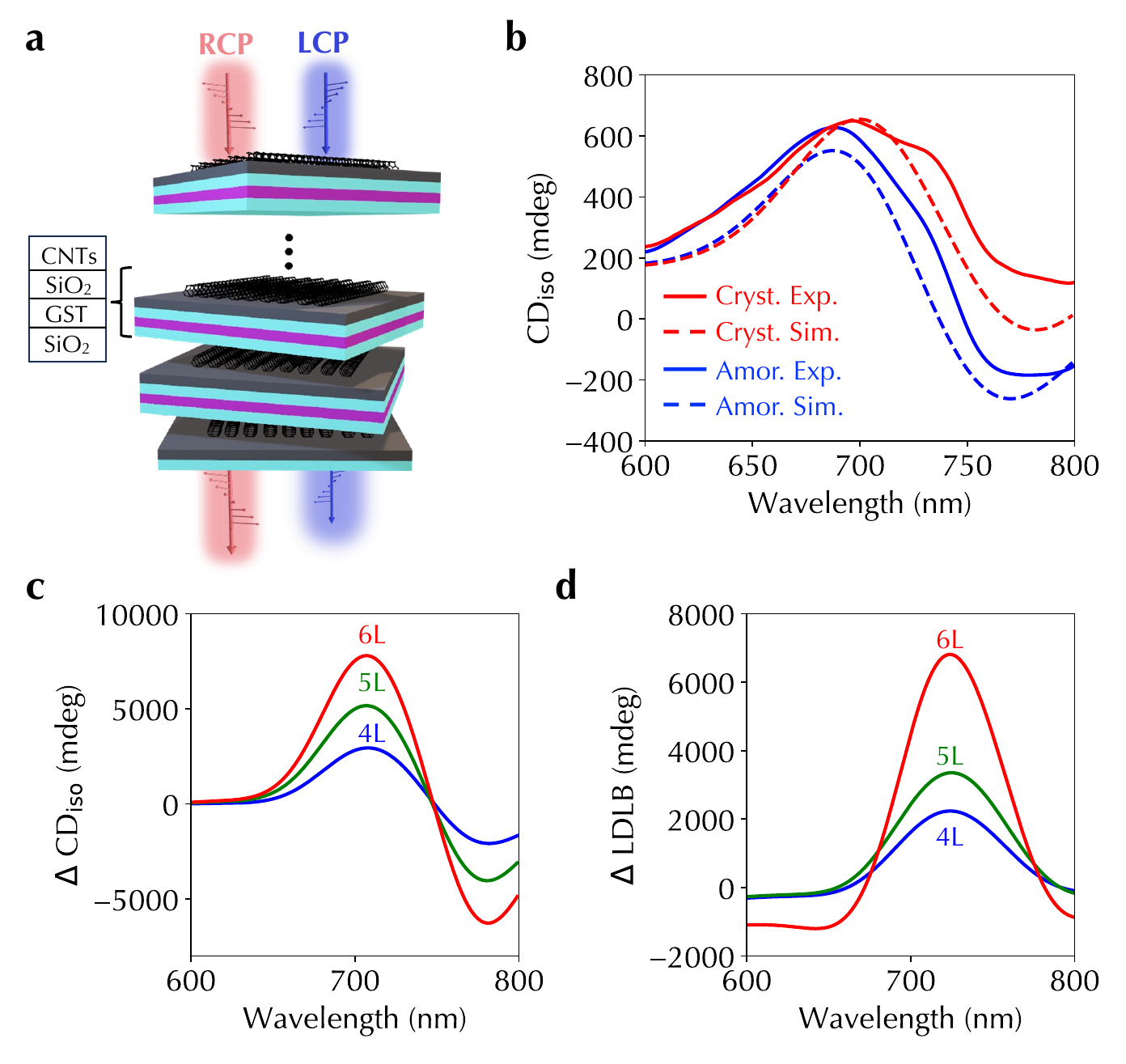}
    \caption{\textbf{Layer scalability of chiroptical responses in the programmable heterostructure}. (a)~Illustration of heterostructures containing multiple CNTs-SiO$_{2}$-GST-SiO$_{2}$ units. (b)~Experimentally measured (solid lines) and simulation (dashed lines) CD$_\textrm{iso}$ spectra under GST amorphous (blue lines) and crystalline (red lines) phases, respectively. (c)~$\Delta$CD$_\textrm{iso}$ and (d)~$\Delta$LDLB spectra for heterostructures containing four (blue lines), five (green lines), and six layers (red lines) of twisted aligned CNT films.}
    \label{fig:fig4}
\end{figure}

\subsection*{Layer scalability}

Further, we demonstrated that both the simulation framework and experimental implementation are scalable with respect to the number of layers in the heterostructure. Specifically, we repeated CNTs-SiO$_{2}$-GST-SiO$_{2}$ configuration in the heterostructure to increase the layer number, as illustrated in Fig.~\ref{fig:fig4}a. We followed the same fabrication processes including the transfer of aligned CNT films and the sputtering deposition of other films; see Methods for more details. Figure~\ref{fig:fig4}b shows experimentally measured CD$_\mathrm{iso}$ spectra under two GST phases (solid lines), which agree with simulation CD$_\mathrm{iso}$ spectra (dashed lines). The maximum $\Delta$CD$_\textrm{iso}$ was 363\,mdeg at 752\,nm wavelength, which is nearly three times as large as that of the heterostructure with two aligned CNT films in Fig.~\ref{fig:fig3}d. In addition, we designed heterostructures containing four, five, and six layers of twisted aligned CNT films for optimizing $\Delta$CD$_\textrm{iso}$ and $\Delta$LDLB. Note that the total number of layers in the heterostructure containing six aligned CNT films is 21. Figure~\ref{fig:fig4}c and \ref{fig:fig4}d show their corresponding spectra. The obtained maximum $\Delta$ CD$_\textrm{iso}$ and $\Delta$LDLB for the six-CNT-layer heterostructure were 7.8\,deg and 6.8\,deg, respectively. The $\Delta$ CD$_\textrm{iso}$ and $\Delta$LDLB spectra under two GST phases are shown in Supplementary Fig.~11 and Fig.~12.

\subsection*{Electrical tunability}

\begin{figure}[hbt]
    \centering
    \includegraphics[width=1.0\textwidth]{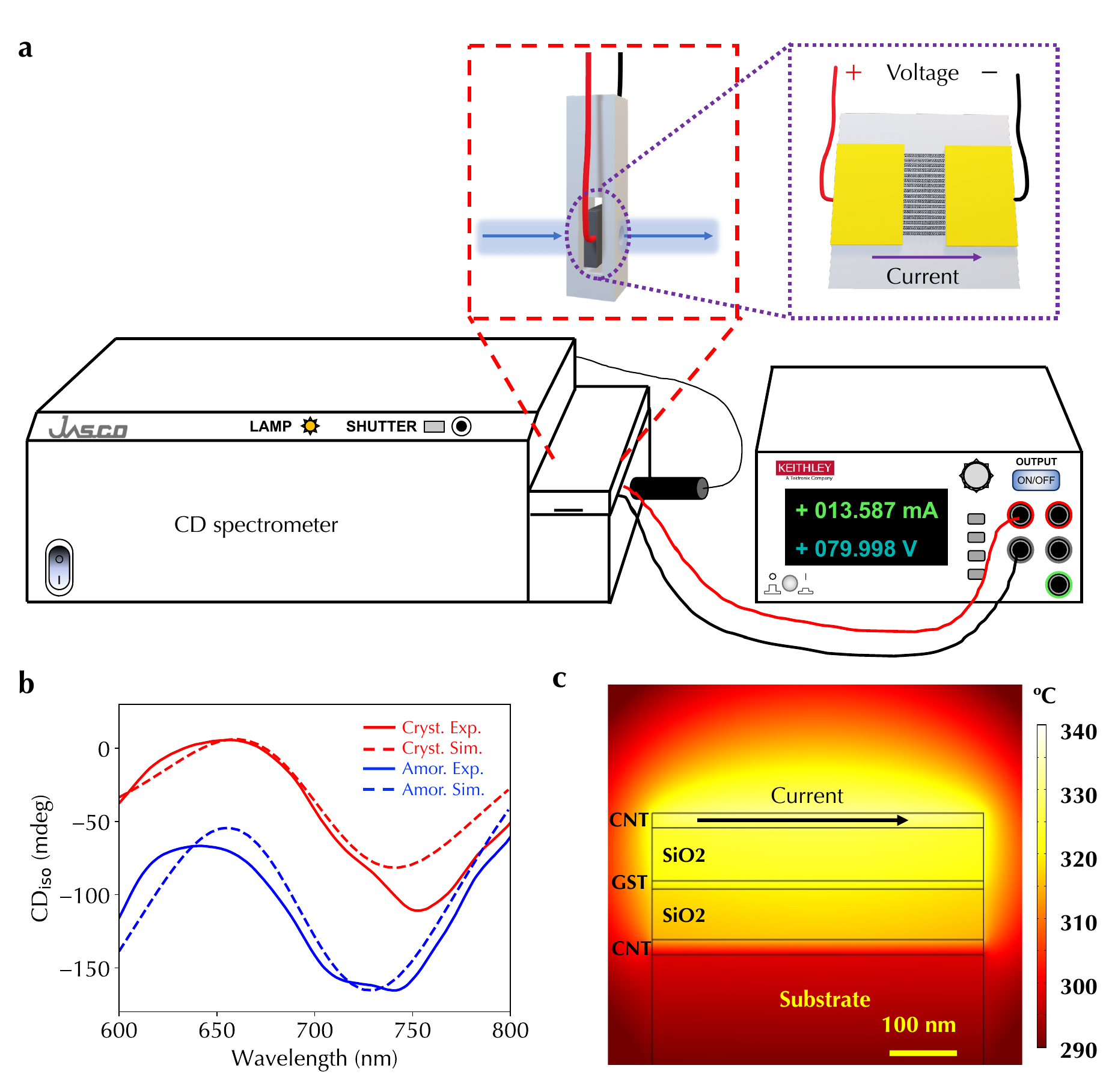}
    \caption{\textbf{Electrical tunability of chiroptical responses in the heterostructure.}. (a)~Illustration of \emph{in-situ} programming of the heterostructure in the CD spectrometer. (b)~Experimentally measured (solid lines) and simulation (dashed lines) CD$_\textrm{iso}$ spectra under GST amorphous (blue lines) and crystalline (red lines) phases, respectively. (c)~Spatial temperature distribution of a scaled-down heterostructure.}
    \label{fig:fig5}
\end{figure}

In addition to contributing to chiroptical responses through twisted stacks, the high thermal conductivity, low heat capacity, and reliable and high current carrying capacity of CNTs~\cite{DresselhausEtAl2001} along the tube axis make aligned CNT films a promising candidate for an efficient Joule heating electrode for programming GST phases. Hence, the aligned CNT film in the heterostructure can play dual roles as both chiroptical and electrode materials. Figure~\ref{fig:fig5}a illustrates \emph{in-situ} programming of the heterostructure in the CD spectrometer. Two electrodes and wires were in contact with the top layer of the aligned CNT film, with the current flowing along the alignment direction. The sample was loaded into the developed 3D-printed sample holder compatible with the spectrometer. We first measured CD spectra under four sample configurations, then applied voltages across the sample through a sourcemeter, and measured CD spectra again under four sample configurations. Figure~\ref{fig:fig5}b displayed measured CD$_\textrm{iso}$ spectra (solid lines), which agreed with simulation spectra (dashed lines) as well. The applied voltage and current for converting the GST film from amorphous to crystalline phases were $\sim80\,$V and $\sim13.6\,$mA, respectively. We further performed multiphysics simulations using the COMSOL Multiphysics software to estimate the temperature profile; see Methods for more details. Figure~\ref{fig:fig5}c shows a spatial distribution of temperature in the heterostructure with scaled-down lateral dimension and delivered electrical energy from the top aligned CNT film. The achieved temperature inside the GST film can be $\sim320^{\circ}$C, which is much larger than the GST transition temperature from amorphous to crystalline phase ($\sim250^{\circ}$C), confirming the phase transition of GST film.

\section*{Discussion}

We have developed a full stack of a versatile experimental platform of programmable chiroptical heterostructure of twisted aligned CNTs and PCMs and corresponding design software infrastructure based on machine learning frameworks including differentiable programming and derivative-free Bayesian optimization. The heterostructure is scalable not only in lateral dimensions to a wafer scale but also in the vertical thickness dimension to a large number of stack layers. We have demonstrated the design and implementation of heterostructures to dynamically program both reciprocal and nonreciprocal CD responses, leading to the discovery of novel phenomena and the development of high-performance devices. Further, aligned CNT films are multifunctional and also serve as electrodes for heating and reconfiguring PCMs. Although the required switching power is large due to the wafer-scale size of the heterostructure, the creation of microscopic nanophotonic planar structures of the heterostructure can substantially reduce switching power, allow reliable multilevel switching, improve cyclability, enable further control of light-matter interaction, and enrich device functionalities. Further, the demonstrated 1D CNT and GST materials in the heterostructure can be expanded to a large library of 1D nanomaterials and electro-optic materials. 

\section*{Methods}

\smallskip
\noindent\textbf{Shaking-assisted vacuum filtration (SAVF)} -- The CNT powder used was purchased from Carbon Solutions, Inc. (product number P2, with a purity $>$90\,wt\%) and synthesized using the arc-discharge method. To prepare aqueous dispersions, 8\,mg of CNT powder was mixed with 20\,mL of an aqueous $0.5\%$ w/w sodium deoxycholate solution, and then was sonicated using an ultrasonic tip horn sonicator (QSonica Q125) for 45 minutes at an output power of 21\,W. Afterward, the sonicated suspension was purified by ultracentrifugation at $247,104\times g$ for two hours to remove large bundles. The supernatant was collected and diluted 27 times before undergoing vacuum filtration using a 1-inch filtration system (MilliporeSigma) with 100-nm-pore-size filter membranes (Whatman Nuclepore Track-Etched polycarbonate hydrophilic membranes, MilliporeSigma). The whole filtration system was placed on a linear shaker (Scilogex SCI-L180-Pro LCD digital linear shaker) as illustrated in Supplementary Fig.~1. Instead of a conventional cylindrical funnel, a funnel with a square-shaped cross-section was fabricated using a 3D printer (Elegoo Saturn 3D Printer). The lateral dimension of the square was 1\,cm. Hence, the linear shaking direction was found to be the alignment direction of obtained aligned CNT films. The linear shaker shook the filtration system at 200\,RPM during the first 15\,minutes of the filtration process. Afterward, the standard filtration process continued without shaking. Close to the end of filtration, the vacuum pump was turned on to fix the alignment structure on the membrane. More details can be found in Ref.~\cite{HeEtAl2016NN}. Supplementary Fig.~1 shows a photo and scanning electron microscopy image of the obtained film, confirming a good alignment along the shaking direction. The obtained film can be transferred onto the desired substrate for characterization and heterostructure fabrication using a wet transfer method. Specifically, a small droplet of water was first placed on the target substrate. The CNT film on the polycarbonate filter membrane was placed with the CNT side in contact with the wet substrate and the polycarbonate side on top. Once the water between the CNT film and the substrate evaporated, the top polycarbonate layer was removed by immersing the sample in a chloroform solution. The sample was finally cleaned with isopropanol. 


\smallskip
\noindent\textbf{TMM calculations} -- A $4\times4$ transfer matrix method was developed to calculate chiroptical responses of the heterostructure containing anisotropic non-magnetic materials under normal incidence. The details can be found in Ref.~\cite{HaoEtAl2008PR} and our prior work~\cite{DoumaniEtAl2023NC}. Briefly, as illustrated in Supplementary Fig.~2, a layer of an isotropic material was modeled with a scalar dielectric function $\varepsilon$ and a thickness $t$. A layer of anisotropic material was modeled with a $2\times2$ dielectric function tensor {$\boldsymbol{\varepsilon}$}, a twist angle $\theta$, and a thickness. Specifically, for an anisotropic material with orthogonal principal axes, such as the directions parallel and perpendicular to CNT alignment, there are four eigenmodes, which are $s$-wave forward, $s$-wave backward, $p$-wave forward, and $p$-wave backward modes, respectively. In the definition of the coordinate systems, waves propagate along the $z$-axis and $x$ and $y$ axes are in the heterostructure plane. $\hat{\textbf{x}}$, $\hat{\textbf{y}}$, and $\hat{\textbf{z}}$ are defined as unit vectors along $x$-, $y$-, and $z$-axes, respectively. The transmitted field $(E_{t,s}, E_{t,p})$, reflected field $(E_{r,s}, E_{r,p})$, and the incident field $(E_{i,s}, E_{i,p})$ can be related through
\begin{equation}
    \left(\begin{array}{c}
        E_{t,s}\\ 0 \\E_{t,p}\\ 0 
    \end{array}\right)
    =Q\left(\begin{array}{c}
        E_{i,s}\\ E_{r,s} \\E_{i,p} \\ E_{r,p}
    \end{array}\right)
    =\left(\begin{array}{cccc}
        Q_{11} & Q_{12} & Q_{13} & Q_{14} \\
        Q_{21} & Q_{22} & Q_{23} & Q_{24} \\
        Q_{31} & Q_{32} & Q_{33} & Q_{34} \\
        Q_{41} & Q_{42} & Q_{43} & Q_{44} \\    
    \end{array}\right)\left(\begin{array}{c}
        E_{i,s}\\ E_{r,s} \\E_{i,p} \\ E_{r,p}
    \end{array}\right),
    \label{eq:tmm_q}
\end{equation}
where $Q$ can be written as a product of a series of $D$ and $P$ matrices. Specifically, for a stack with $N$ layers, 
\begin{equation}
    Q=D_{N+1}^{-1}D_{N}P_{N}D_{N}^{-1}D_{N-1}P_{N-1}D_{N-1}^{-1}\cdot\cdot\cdot D_{1}P_{1}D_{1}^{-1}D_{0}
    \label{eq:q_exp}
\end{equation}
    with
\begin{equation}
    {D_j}=\left(\begin{array}{cccc}
    \textrm{sin}\theta_j & \textrm{sin}\theta_j & \textrm{cos}\theta_j & \textrm{cos}\theta_j \\
    -n_{s_j}\textrm{sin}\theta_j & n_{s_j}\textrm{sin}\theta_j & -n_{p_j}\textrm{cos}\theta_j & n_{p_j}\textrm{cos}\theta_j \\
    n_{s_j}\textrm{cos}\theta_j & -n_{s_j}\textrm{cos}\theta_j & -n_{p_j}\textrm{sin}\theta_j & n_{p_j}\textrm{sin}\theta_j \\
    \textrm{cos}\theta_j & \textrm{cos}\theta_j & -\textrm{sin}\theta_j & -\textrm{sin}\theta_j \\
    \end{array}\right)
\end{equation} 
    and
\begin{equation}
    {P_j}=\left(\begin{array}{cccc}
    e^{\textrm{i}k_{j,1}t_j} & 0 & 0 & 0\\
    0 & e^{\textrm{i}k_{j,2}t_j} & 0 & 0\\
    0 & 0 &  e^{\textrm{i}k_{j,3}t_j} & 0\\
    0 & 0 & 0 &  e^{\textrm{i}k_{j,4}t_j}
    \end{array}\right),
\end{equation}
for $j$-th layer and $j\in[0,N+1]$. Here, for the $j$-th layer, the $t_j$ is the thickness, $k_{j,1} = -k_{j,2} = k_0n_{s_j}$, $k_{j,3} = -k_{j,4} = k_0n_{p_j}$, $n_{s_j}$ is the refractive index along the $s_j$ axis, $n_{p_j}$ is the refractive index along the $p_j$ axis, $\theta_j$ is the angle of the $s_j$ axis with respect to the $x$-axis in counterclockwise rotation, and $k_0$ is vacuum wavevector. For layers with isotropic materials, such as the input and output layers, the principal axes are chosen to be $s_1 = s_{N+1} = \hat{\textbf{x}}$ and $p_1 = p_{N+1} = \hat{\textbf{y}}$. For layers with anisotropic materials, the dielectric function tensor {$\boldsymbol{\varepsilon}_j$} in $xy$ coordinate systems can be written as 
\begin{align}
    \boldsymbol{\varepsilon}_j = 
    R_j\left(\begin{array}{cc}
        \varepsilon_{s_j} & 0 \\0 & \varepsilon_{p_j}
    \end{array}\right){R_j}^{-1}, \\
    {R_j} = \left(\begin{array}{cc}
        \textrm{cos}\theta_j & -\textrm{sin}\theta_j \\\textrm{sin}\theta_j & \textrm{cos}\theta_j
    \end{array}\right),
\end{align}
with $n_{s_j}=\sqrt{\varepsilon_{s_j}}$ and $n_{p_j}=\sqrt{\varepsilon_{p_j}}$. As a result, we can obtain transmission and reflection coefficients in terms of the matrix elements of $Q$ as

\begin{equation}
    r_{ss} = \left.\frac{E_{r,s}}{E_{i,s}}\right\vert_{E_{i,p} = 0} 
    = \frac{Q_{24}Q_{41} - Q_{21}Q_{44}}{Q_{22}Q_{44} - Q_{24}Q_{42}},
\end{equation}

\begin{equation}
    r_{sp} = \left.\frac{E_{r,p}}{E_{i,s}}\right\vert_{E_{i,p} = 0} 
    = \frac{Q_{21}Q_{42} - Q_{22}Q_{41}}{Q_{22}Q_{44} - Q_{24}Q_{42}},
\end{equation}

\begin{equation}
    t_{ss} = \left.\frac{E_{t,s}}{E_{i,s}}\right\vert_{E_{i,p} = 0} = Q_{11}+\frac{Q_{12}(Q_{24}Q_{41}-Q_{21}Q_{44})+Q_{14}(Q_{21}Q_{42}-Q_{22}Q_{41})}{Q_{22}Q_{44}-Q_{24}Q_{42}}
\end{equation}

\begin{equation}
    t_{sp} = \left.\frac{E_{t,p}}{E_{i,s}}\right\vert_{E_{i,p} = 0} = Q_{31}+\frac{Q_{32}(Q_{24}Q_{41}-Q_{21}Q_{44})+Q_{34}(Q_{21}Q_{42}-Q_{22}Q_{41})}{Q_{22}Q_{44}-Q_{24}Q_{42}}\ .
\end{equation}


\begin{equation}
    r_{ps} = \left.\frac{E_{r,s}}{E_{i,p}}\right\vert_{E_{i,s} = 0} 
    = \frac{Q_{24}Q_{43} - Q_{23}Q_{44}}{Q_{22}Q_{44} - Q_{24}Q_{42}},
\end{equation}

\begin{equation}
    r_{pp} = \left.\frac{E_{r,p}}{E_{i,p}}\right\vert_{E_{i,s} = 0} 
    = \frac{Q_{23}Q_{42} - Q_{22}Q_{43}}{Q_{22}Q_{44} - Q_{24}Q_{42}},
\end{equation}

\begin{equation}
    t_{pp} = \left.\frac{E_{t,s}}{E_{i,p}}\right\vert_{E_{i,s} = 0} = Q_{33}+\frac{Q_{32}(Q_{24}Q_{43}-Q_{23}Q_{44})+Q_{34}(Q_{23}Q_{42}-Q_{22}Q_{43})}{Q_{22}Q_{44}-Q_{24}Q_{42}},
\end{equation}

\begin{equation}
    t_{ps} = \left.\frac{E_{t,p}}{E_{i,p}}\right\vert_{E_{i,s} = 0} = Q_{13}+\frac{Q_{12}(Q_{24}Q_{43}-Q_{23}Q_{44})+Q_{14}(Q_{23}Q_{42}-Q_{22}Q_{43})}{Q_{22}Q_{44}-Q_{24}Q_{42}}.
\end{equation}

The input light of any other polarization states can be represented as a linear combination of $E_{i,s}$ and $E_{i,p}$. For example, LCP light can be represented as $0.5E_{i,s}+ 0.5iE_{i,p}$, RCP light can be represented as $0.5E_{i,s}- 0.5iE_{i,p}$. Any linearly polarized with the angle between polarization direction and $s$-wave polarization direction, $\alpha$, can be represented as cos$(\alpha)E_{i,s}$ +  sin$(\alpha)E_{i,p}$. Hence, the output fields and their amplitude and phase can be calculated using the input field vector in Eq.~\ref{eq:tmm_q}. Since the operations described above are all matrix-matrix multiplications, \texttt{PyTorch} (version 1.9.0) was used to implement the TMM calculation framework with a built-in backpropagation algorithm for gradient descent-based optimization. A Nvidia GeForce RTX 3090 GPU card with a cuda version 11.1 was used to accelerate calculations. 

\smallskip
\noindent\textbf{Four-configuration CD simulation and measurement} -- Theoretical analyses~\cite{AlbanoEtAl2020CR,AlbanoEtAl2022C} have shown that the observed CD signals from solid-state samples (CD$_\textrm{obs}$) can be described as CD$_\textrm{obs}$ = CD$_\textrm{iso}$ + 0.5(LD$^{'}$LB $-$  LDLB$^{'}$) + $\alpha_{\mathrm{c}}$(LD$^{'}$sin($2\theta$) - LDcos($2\theta$)). The first term CD$_\textrm{iso}$ is the intrinsic component of CD, which is from excitonic transitions in CNTs and is isotropic and independent of sample orientation. The second LDLB term is from the interference of the sample's LD and LB responses. This term is not from excitons, independent of instrumental faults, and not an artifact. It represents a real, perfectly reproducible differential attenuation of LCP and RCP light. This LDLB term is invariant upon sample rotation around the axis perpendicular to the sample plane but inverts the sign under sample flipping. The third term artifact comes from the coupling of residual birefringence in the instrument and LD and LB of samples. This artifact term inverts the sign under $90^{\circ}$ sample rotation around the axis perpendicular to the sample plane. Based on these properties, a four-configuration protocol was employed to extract the CD$_\textrm{iso}$ and LDLB terms. Specifically, when the sample was positioned as illustrated in Supplementary Fig.~3, four CD spectra were calculated or measured and denoted as CD$_{\mathrm{m1}}$, CD$_{\mathrm{m2}}$, CD$_{\mathrm{m3}}$, and CD$_{\mathrm{m4}}$. The average spectra under in-plane rotation, which are 0.5(CD$_{\mathrm{m1}}$ + CD$_{\mathrm{m2}}$) or 0.5(CD$_{\mathrm{m3}}$ + CD$_{\mathrm{m4}}$), can remove the third term. Hence, the term 0.5(CD$_{\mathrm{m1}}$ + CD$_{\mathrm{m2}}$) (i.e., the observed CD, CD$_\mathrm{obs}$, from the front side) corresponds to CD$_\textrm{iso}$ + 0.5(LD$^{'}$LB $-$  LDLB$^{'}$) and the term 0.5(CD$_{\mathrm{m3}}$ + CD$_{\mathrm{m4}}$) (i.e., the observed CD, CD$_\mathrm{obs}$, from the back side) corresponds to CD$_\textrm{iso}-$  0.5(LD$^{'}$LB $-$  LDLB$^{'}$). Hence, CD$_\textrm{iso}$ = 0.25(CD$_{\mathrm{m1}}$ + CD$_{\mathrm{m2}}$ + CD$_{\mathrm{m3}}$ + CD$_{\mathrm{m4}}$) and 0.5(LD$^{'}$LB $-$  LDLB$^{'}$) = 0.25(CD$_{\mathrm{m1}}$ + CD$_{\mathrm{m2}} -$ CD$_{\mathrm{m3}} -$ CD$_{\mathrm{m4}}$). 

\smallskip
\noindent\textbf{Linear- and circular-polarization dependent optical spectroscopy} -- CD spectra and average absorption spectra of LCP and RCP light were measured using a Jasco J-810 CD spectrometer, covering a wavelength range of $200-900$\,nm. Linearly polarized absorption spectra were measured using an ultraviolet–visible-near-infrared (UV–vis-NIR) spectrometer (Perkin Elmer Lambda 950 UV–vis-NIR) equipped with an automatically controlled rotating broadband polarizer in the same wavelength range. The incident beam with a diameter of 2\,mm was defined by a customized 3D-printed sample holder, and was the same in all measurements. In the design and fitting of heterostructures, the thicknesses of aligned CNT films were benchmarked by comparing the peak average absorption of LCP and RCP light in the UV range with that of a reference sample whose thickness was measured using an atomic force microscope (Parksystems NX20)~\cite{DoumaniEtAl2023NC}. The peak average UV absorption was assumed to be proportional to the sample thickness. 

\smallskip
\noindent\textbf{Measurement of dielectric functions of aligned CNTs} -- The dielectric functions parallel and perpendicular to the CNT alignment direction from the UV to NIR ranges were modeled as a summation of Voigt functions~\cite{MenesesEtAl2005JNS}

\begin{equation}
    \epsilon_{s,p}(\omega) = \epsilon_{\infty,s,p} + \sum_{n=1}^{N}\left.C_{\textrm{V},s,p}(\omega)\right\vert_{A_{n}, \omega_{0,n}, \gamma_{\textrm{L},n}, \gamma_{\textrm{G},n}}
\end{equation}
with 
\begin{align}
    \left.C_\textrm{V}(\omega)\right\vert_{A, \omega_{0}, \gamma_\textrm{L}, \gamma_\textrm{G}} &= -A\frac{\textrm{Im}(F(x - x_0 - \textrm{i}y) + F(x + x_0 + \textrm{i}y))}{\textrm{Re}(F(\textrm{i}y))}\\
    &+ \textrm{i}A \frac{\textrm{Re}(F(x - x_0 + \textrm{i}y) - F(x + x_0 + \textrm{i}y))}{\textrm{Re}(F(\textrm{i}y))}
\end{align}
and 
\begin{equation}
    x = \frac{2\sqrt{\textrm{ln}2}}{\gamma_\textrm{G}}\omega, x_0 = \frac{2\sqrt{\textrm{ln}2}}{\gamma_\textrm{G}}\omega_0, y = \frac{\gamma_\textrm{L}\sqrt{\textrm{ln}2}}{\gamma_\textrm{G}},
\end{equation}
where $\omega$ is angular frequency, $A$ is amplitude factor, $\omega_0$ is resonance frequency, $\gamma_\textrm{L}$ is Lorentz linewidth, $\gamma_\textrm{G}$ is Gaussian linewidth, and $F$ is Faddeeva function. $N$ was selected as 5. Hence, for each polarization ($s$ or $p$), there were 21 fitting parameters, including \{$A_{n}, \omega_{0,n},\gamma_{\textrm{L},n},\gamma_{\textrm{G},n}$, $n=1-5$\} and $\epsilon_{\infty}$, and 42 fitting parameters in total for both polarizations. These fitting parameters were uniquely determined by simultaneously fit six experimentally measured spectra, including linear-polarization absorption spectra for one-layer aligned CNT film, and averaged absorption of LCP and RCP light and CD spectra of twisted two-layer and three-layer stacks with a twist angle of 30$^{\circ}$, as shown in Supplementary Fig.~4. The fitting wavelength range was from 600\,nm to 800\,nm wavelength.

\smallskip
\noindent\textbf{Measurement of dielectric functions of GST films} -- The film thickness and optical constants (refractive index $n$ and extinction coefficient $k$) of GST films were measured using a J.A. Woollam RC2 Spectroscopic Ellipsometer over a wavelength range of $190-1700$\,nm. The measurements were conducted over a wide spectral range from UV to NIR and were analyzed using CompleteEASE software. In this procedure, polarized light was reflected off the sample surface, and the change in polarization was measured as two quantities: $\Psi$ and $\Delta$. $\Psi$ represents the amplitude ratio and $\Delta$ represents the phase difference. A model describing the sample's physical structure (layers and materials) was created and adjusted to fit the measured $\Psi$ and $\Delta$ values through regression analysis. The software calculated results based on the model, and the model parameters (e.g., layer thicknesses, optical constants) were varied to minimize the mean-squared error between experimental and calculation results. The fitting process yielded the film thickness and the optical constants of the material. These optical constants include the real and imaginary parts of refractive indices, or alternatively, the real and imaginary parts of the dielectric function $\varepsilon$. 


\smallskip
\noindent\textbf{Backpropagation and Bayesian optimization algorithms} -- The Adam optimizer in \texttt{PyTorch} was employed for the backpropagation algorithm. The learning rate was set at 0.01. For Bayesian optimization, the results shown in Fig.~\ref{fig:fig3}c were obtained without the bound of the twist angle between aligned CNT films. For the results shown in Supplementary Fig.~5, the twist angle was bounded in a range of $22.5-67.5^{\circ}$. The accumulated average over all epochs was plotted. 

\smallskip
\noindent\textbf{Heterostructure fabrication} -- The heterostructure fabrication process consists of (i)~transfer of aligned CNT films, (ii)~deposition of GST and other dielectrics, and (iii)~twist-stacking of aligned CNT films. The substrate was fused silica for broadband transparency. For (i), aligned CNT films were produced through the SAVF method as described before. The alignment direction was determined by the linear shaking direction. The produced film was typically cut into multiple pieces for twist-stacking and transferred onto the substrate or heterostructure using the wet transfer process as described before. For (ii), SiO$_{2}$ films were deposited using a Denton Discovery 18 Sputtering System at an argon pressure of 6\,mTorr with a power setting of 100\,W. GST films were deposited using the same system at an argon pressure of 4.5\,mTorr and a power setting of 35\,W. For (iii), to facilitate twist-stacking, the heterostructure was put onto a transparent protractor, which was back-illuminated by a light-emitting-diode panel. An aligned CNT film was placed on top of the heterostructure and the orientation was rotated to a specific angle relative to transferred aligned CNT films before. The film was then transferred using the same wet transfer process. To induce the phase transition in GST films, the heterostructure was placed on a hotplate (Corning hotplate and stirrer with digital display) at set temperatures and left for a few minutes for a complete phase transition. 

\smallskip
\noindent\textbf{Finite element simulation} -- A 2D finite element simulation using COMSOL Multiphysics was conducted to analyze the temperature distribution of the heterostructure under applied voltage and current. Because of the limited computer memory resource and wafer-scale sample sizes, the finite element model and thermal excitation were simultaneously scaled down. Specifically, the thickness or vertical dimension of the heterostructure was kept unchanged, while the lateral dimension was scaled down to $500 \times 500$ nm$^2$. The thicknesses of the layers from top to bottom were 20\,nm for the top CNT film, 71\,nm for SiO$_2$, 11\,nm for GST, 68\,nm for SiO$_2$, and 20\,nm for the bottom CNT film. The thermal conductivity of CNTs along the tube axis was set as 43\,W m$^{-1}$ K$^{-1}$~\cite{YamaguchiEtAl2019APL}, and the GST thermal conductivity was set as 0.27\,W m$^{-1}$ K$^{-1}$~\cite{TaghinejadEtAl2021OE}. The heat source was a square wave with a duration of 5\,$\mu$s and a power of $3\times10^{-5}$\,W, resulting in a total heat energy of $1.5\times 10^{-10}$\,J. This was scaled down from the injected heat energy into our fabricated device with a lateral size of of 5\,mm$\times$5\,mm, which was estimated with a power $1.1$\,W and phase transition time $\sim1-2$\,ms. 

\section*{Data availability}
The data that support the findings of this study are available from the corresponding author upon request. 

\section*{Acknowledgements}
J.F., R.C., H.X., and W.G.\ acknowledge support from the National Science Foundation through Grants No.\ 2230727, No.\ 2235276, No.\ 2316627 and No.\ 2321366. 

\section*{Author Contributions Statement}
W.G.\ conceived the idea, designed experiments, and supervised the project. J.F.\ performed the experiments with the help of R.C.\ and H.X.\ and under the support and guidance W.G. J.F.\ and M.L.\ conducted theoretical modeling and calculations with the help of Y.T.\ and under the support and guidance of W.G. N.H.\ performed spectroscopic ellipsometry measurements. All authors discussed the results and contributed to the manuscript. 

\section*{Competing Interests Statement}
The authors declare no competing interests.



\end{document}